\documentclass[useAMS,usenatbib]{mn2e}
\newcommand{\HI}{{\sc H\,i}}
\newcommand{\mJybeam}{mJy beam$^{-1}$}
\newcommand{\msun}{{$M_\odot$}}
\newcommand{\kms}{$\,$km$\,$s$^{-1}$}
\newcommand{\ltsima} {$\; \buildrel < \over \sim \;$}
\newcommand{\gtsima} {$\; \buildrel > \over \sim \;$}
\newcommand{\lta} {\lower.5ex\hbox{\ltsima}}
\newcommand{\gta} {\lower.5ex\hbox{\gtsima}}

\newcommand{\pc}{pc$^{-2}$}

\newcommand{\ergs}{$\,$erg$\,$s$^{-1}$}

\usepackage{txfonts}
\usepackage{graphicx}
\usepackage{url}

\usepackage{aas_macros}
\usepackage{threeparttable}
\usepackage{dcolumn}
\title[The complex H{\,\small I} structure of NGC~3079]{The ``shook up" galaxy NGC~3079:  the complex interplay between H{\,\LARGE I}, activity and environment}
\author[N.Shafi et al.]{N. Shafi$^{1,2}$, T.A. Oosterloo$^{3,4}$, R. Morganti$^{3,4}$,  S. Colafrancesco$^{1}$, R. Booth$^{5}$\\\
$^1$ University of the Witwatersrand, Private Bag X03, Wits 2050, South Africa\\
$^2$ Hartebeesthoek Radio Astronomy Observatory, Krugersdorp 1740, South Africa\\
$^3$ ASTRON - the Netherlands Institute for Radio Astronomy, Postbus 2, 7990 AA Dwingeloo, The Netherlands \\
$^4$ Kapteyn Astronomical Institute, University of Groningen, Postbus 800, 9700 AV Groningen, The Netherlands \\
$^5$ University of Pretoria, Private bag X20, Hatfield 0028, South Africa\\
}

\date{}

\pagerange{\pageref{firstpage}--\pageref{lastpage}}
\pubyear{}
\def\LaTeX{L\kern-.36em\raise.3ex\hbox{a}\kern-.15em
    T\kern-.1667em\lower.7ex\hbox{E}\kern-.125emX}

\begin{document}

\maketitle

\label{firstpage}

\begin{abstract}
We present deep neutral hydrogen (\HI) observations of the starburst/Seyfert galaxy NGC~3079 and its environment, obtained with the Westerbork Synthesis Radio Telescope. Our observations reveal previously unknown components, both in \HI\ emission and in absorption, that show that NGC~3079 is going through a hectic phase in its evolution.
The \HI\ disk appears much more extended than previously observed and is morphologically and kinematically lopsided on all scales with evidence for strong non-circular motions in the central regions. Our data reveal prominent gas streams encircling the entire galaxy suggesting strong interaction with its neighbours. A 33-kpc long \HI\ bridge is detected between NGC~3079 and MCG~9-17-9, likely caused by ram-pressure stripping of MGC~9-17-9 by the halo of hot gas of NGC~3079. The cometary \HI\ tail of the companion NGC~3073, earlier discovered by Irwin et al., extends about twice as long in our data, while a shorter, second tail is also found. This tail is likely caused by ram-pressure stripping by the strong, starburst driven wind coming from NGC~3079. We also detect, in absorption,  a nuclear \HI\ outflow extending to velocities well outside what  expected for gravitational motion. This is likely an atomic counterpart of the well-studied outflow of ionised gas present in this galaxy. This may indicate that also large amounts of {\sl cold} gas are  blown out of NGC~3079 by the starburst/AGN. Our estimates of the jet energy and kinetic power suggest that both the AGN and the starburst in NGC~3079 are powerful enough to drive the atomic outflow.
\end{abstract}


\section{Introduction}\label{s:intro}

Deep observations of 21-cm atomic neutral hydrogen (\HI) are known to provide a powerful tracer of  the structure and evolution of galaxies. Given that in general the \HI\ distribution is often more extended than the stellar body of a galaxy, it provides an excellent, and sometimes unique,  diagnostic of  the role of the environment and  interactions in shaping galaxy properties.  It also allows, through observations of outflows of cold gas, to study the interplay between  the ISM and the energetic phenomena associated with  active nuclei \citep[e.g.,][]{Gallimore1999,Morganti2013} and star formation  \citep[e.g.,][]{Hulst2001,Boomsma2005,Walter2008,Koribalski2010,Heald2011,Wang2013}.

Here we present a detailed study of the  \HI\ in NGC~3079 and in its environment. NGC 3079 is well-known for its intense starburst in the central regions and an associated outflow of ionised gas, while it also harbours a Seyfert nucleus. Although it has been the target of many detailed studies, we have obtained new \HI\ data that are much more sensitive than those from previous observations with the aim to do a more detailed study of the role of the environment in the elevated activity in this galaxy and, vice versa, the effect of NGC 3079 on it companion galaxies.

NGC~3079 is a nearby spiral galaxy which has been studied extensively in different wavelengths and resolutions because it is known for its several unique properties. One of the most spectacular features of NGC 3079 is the ionised gas outflow (super-wind)  forming a ``super-bubble" at the eastern side of the galactic plane \citep{Cecil2001}. The outflow appears to originate from the nucleus and reaches velocities up to $\sim$ 1500 \kms \citep{Cecil2001}. Chandra X-ray observations also show a clear correspondence between X-ray emission and the optical H$\alpha$ filaments extending $\sim$ 1.3 kpc from the nucleus \citep{Cecil2002}. 
Large-scale outflows have also been observed in the radio continuum as two, kpc-scale, radio lobes extending on both sides of the major axis of the galaxy \citep{debr77,b18,b14}. 
 NGC~3079 is classified as Seyfert 2 \citep{ford86} or LINER (Low Ionization Narrow Emission Line Region, \citealt{b20}).

The ``super-bubble" can be caused by either an AGN \citep[e.g.,][]{irwin88} or a nuclear starburst \citep[e.g.,][]{sof01}. 
Indeed,  various studies of the nucleus of the galaxy  have provided evidence for the presence of both a weak AGN and a nuclear starburst. Several X-ray observations provide compelling evidence for the existence of an obscured AGN. For instance, the presence of an X-ray continuum excess in the $20 -100$ keV band and a 6.4 keV Fe K$_{\alpha}$ line emission \citep{Cecil2002}, and the detection of both soft and hard X-ray emission showing a high hydrogen column density ($N_{\rm H} \sim 10^{25}$ cm$^{-2}$) towards the nucleus \citep{Iyomoto2001}. The presence of an AGN is also indirectly supported by Very Long Baseline Interferometry (VLBI) observations of  \citet{irwin88} at 5 GHz which revealed aligned multiple-components A, B and C (along P.A.$=120^{\circ}$) consistent with a  pc-scale nuclear radio jet. This was later supported by several high resolution radio continuum observations \citep{b14,trot98,swada00,hagiwa04,kondra05,middel07}. These high resolution observations also revealed additional components (E and F) to the central VLBI source with varying spectral indices. Although the additional components have similar orientation and spacing as the components detected by \citet{irwin88}, they are not aligned with either the kpc-scale radio lobe or the proposed pc-scale jet \citep{kondra05,middel07}. \cite{kondra05} have also suggested that the jet may exist with a pc-scale wide angle outflow, which goes to show that the central source is far more complex than reported in \citet{irwin88}.

Strong FIR emission \citep{soif87}, colour excess \citep{lawerence85}, and unusually high-density core of molecular gas relative to other starburst galaxies or AGN ($\sim$7200 \msun \pc; \citealt{plan97,saka99,sof01,koda02}) all point towards a nuclear starburst (but see also  \citealt[][for counter-arguments]{haward95}). 

Previous VLA \HI\ observations of NGC 3079 by \citet{b32} showed a smooth disk of \HI\ with however some signs of  perturbation, e.g.\ in the southern part.  One of the main aims of our study is to  investigate whether deeper \HI\ observations would provide further evidence for such perturbations and  whether these may be related to the activity observed in NGC 3079. 

\citet{b14} detected \HI\ and hydroxyl (OH) multi-component absorptions towards the centre. The results from these observations suggest a combination of rotation and gas outflow to be present in the nuclear region \citep{b14,b21}. The second aim of our work is, exploiting the high spectral stability of the Westerbork Synthesis Radio Telescope (WSRT), to see whether deeper observations would provide more information about this outflow of cold gas.

NGC 3079 is seen edge-on ($i=84^{\circ}$; \citealt{b20}) and is at a distance of $\sim$ 16 Mpc (i.e.\ 1 arcsec = 85 pc). NGC~3079 has two known nearby companion galaxies, NGC~3073 (dwarf S0) and MCG~9-17-9 (small Sb-Sc). The optical appearances of both companions do not show any disturbance. However in  \HI,  NGC~3073 appears to be affected by the nuclear activity in NGC~3079 \citep{b1}. This is observed as an elongated \HI\ tail coming from NGC~3073 pointing away from the nucleus of NGC~3079 in the data from \citet{b1}. Interestingly, as \citet{b1} argue, this tail may be caused by ram-pressure stripping  of \HI\ from NGC 3073 by the starburst-driven superwind of NGC 3079. Our deeper \HI\ data allow us to study this tail in more detail.

Our observations and the results are presented in Sections \ref{s:reduction} and \ref{s:result} followed by discussion in Sections  4, 5, and 6.

\begin{figure*}
\centering
\includegraphics[width=16cm]{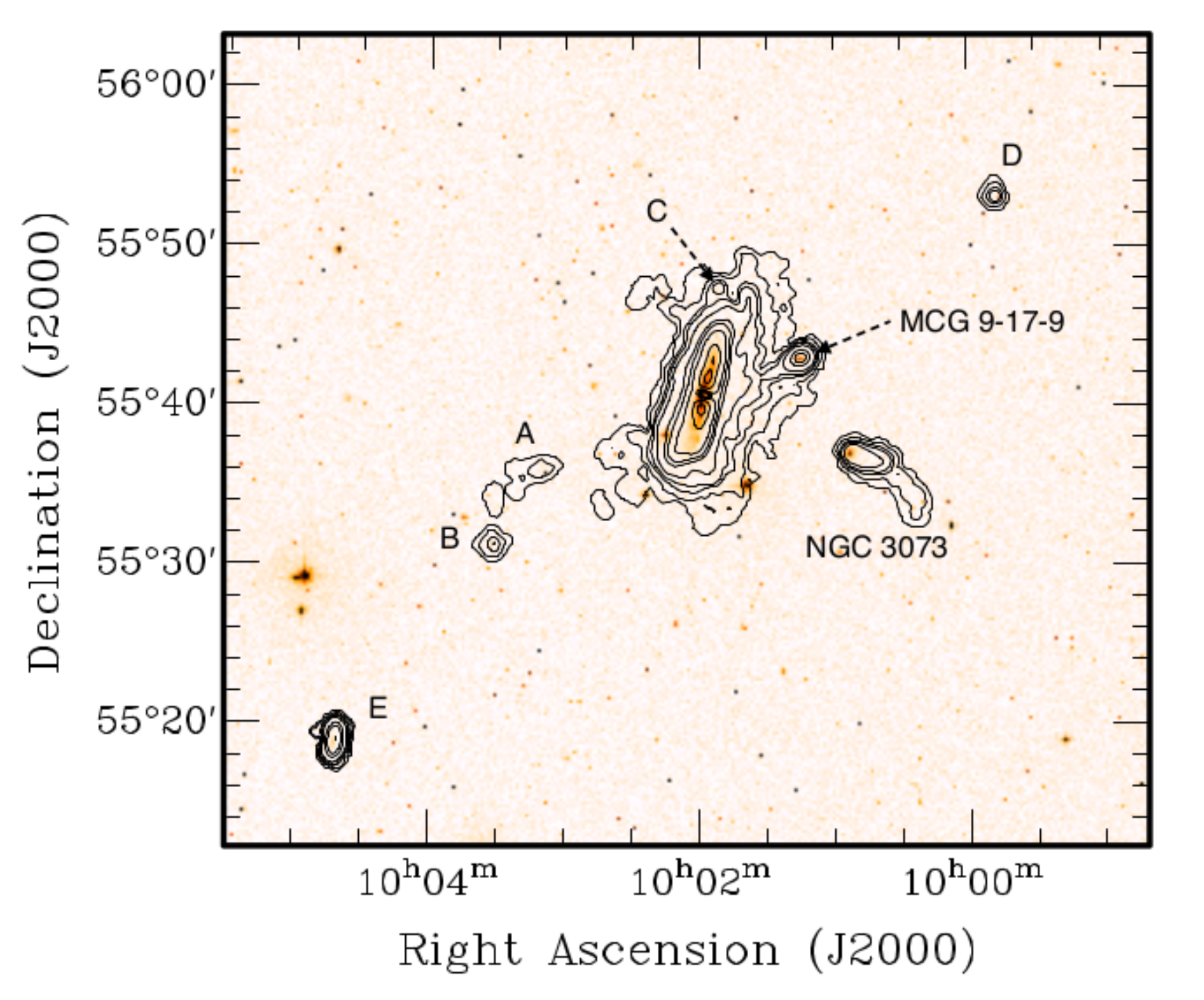}
\caption{Total \HI\ distribution in and around NGC~3079 at $45^{\prime\prime} \times 48^{\prime\prime}$ resolution overlaid onto the optical DSS2 image. Contours are at the levels of 0.05, 0.1, 0.3, 0.5, 1, 3, 5, 10, 30 and 50 $\times 10^{20}$ cm$^{-2}$ . The positions of known and possible companion galaxies are marked, see Table \ref{t:parameter} for details. }
\label{f:totHI2}
\end{figure*}

\section{Observations and Data Reduction}\label{s:reduction}

NGC~3079 was observed with the Westerbork Synthesis Radio Telescope (WSRT). Five 12-hr observations were performed on  4, 5, 9, 12 and 28 Jan 2004. A bandwidth of 20~MHz with 1024 channels was used while the observations used Doppler tracking at a heliocentric velocity of 1135 \kms.  The  parameters of the observations are summarised  in Table \ref{t:obs}.

The data were calibrated and reduced using the {\sc miriad} package \citep{sault95}. Two standard calibrators, 3C~286, and 3C~48, were used to calibrate the flux scale and the spectral response (bandpass). Separate data sets  were generated for line and for continuum.  This was done by fitting a 4-th order polynomial to those channels of each visibility spectrum ('uvlin') that do not contain line emission.

\begin{table}
 \caption{Observational Parameters.}\label{t:obs}
\begin{center}
 \begin{tabular}{l  c}\hline\hline
Pointing centre $\alpha$ (J2000)&10$^{\rm h}$ 01$^{\rm m}$ 57.8$^{\rm s}$\\
Pointing centre $\delta$ (J2000)&\llap{+}55$^\circ$ 40$^\prime$ 47.0$^{\prime\prime}$\\
Date  &    4, 5, 9, 12, 28 Jan 2004 \\
Integration time&$5 \times 12$ hr\\
Doppler tracking velocity&1135 \kms\\
Bandwidth&20 MHz\\
Number of channels&1024\\
Velocity resolution &16.5 \kms\\
\hline
{Line cubes}  &  \\
Synthesised beam ($\alpha\times\delta$)& rms noise \\
$19^{\prime\prime}\times 23^{\prime\prime}$&0.13 \mJybeam\\
$29^{\prime\prime}\times 33^{\prime\prime}$&0.13 \mJybeam\\
$45^{\prime\prime}\times 48^{\prime\prime}$&0.17 \mJybeam\\
$78^{\prime\prime}\times 80^{\prime\prime}$&0.23 \mJybeam\\
\hline
{Continuum image}  &  \\
Synthesised beam ($\alpha\times\delta$)& rms noise \\

$12^{\prime\prime}\times 14^{\prime\prime}$& 0.08 \mJybeam\\
Peak flux  & 310 \mJybeam \\
\hline
\end{tabular}\\
\vspace{0.2cm}
\end{center}
\end{table}

To improve the dynamic range of the images, self-calibration was necessary to correctly determine the antenna gains as a function of time. This was done using the continuum data where iteratively the antenna gains and a model of the continuum sources   was determined, until convergence was achieved. The calibration of the telescope obtained in this way was also copied from the continuum data to the line data. However, by only following this procedure, the quality of the images was still limited by the 
presence of strong off-axis sources that could not be calibrated well due to direction-dependent imaging errors. 
These off-axis errors are  due to 
small pointing/tracking errors and frequency dependent, quasi-periodic variation in the primary beam  due to standing waves in the dishes of the WSRT.
Therefore, we have used the technique commonly known as ``peeling"  to correct for these off-axis errors in the image, following  relatively standard steps \citep[e.g.][]{oosterloo2011}. 
The procedure involves, after subtraction of the central sources from the $uv$ data, self-calibration on the remaining problematic off-axis sources (in this case   including the well-known double quasar and first discovered gravitational lens Q~0957+561 A, B) to derive  extra, direction-dependent corrections. The continuum image improved significantly after peeling as most image artefacts were removed while reaching an rms noise level  of  0.08 \mJybeam\ which is  32\%  lower than  prior to peeling.

For the line data,  cubes were made using different image  weightings and taperings. The data were binned a factor 2 in frequency after which  additional Hanning smoothing was applied resulting in a final velocity resolution of $16.5$ \kms.  Parameters describing the data are listed in Table \ref{t:obs}.

The cubes were cleaned using the Clark algorithm \citep{Clark1980}. In order to be able to clean to low flux levels,  the cleaning was done using masks that were defined by smoothing the data to twice the spatial resolution and by identifying regions with line emission in these smoothed images. The spatial resolution of  the final cubes  ranges between  about  $20^{\prime\prime}$ and $80^{\prime\prime}$ (see Table \ref{t:obs}).  The noise levels, per channel, range between 0.13 and 0.23 \mJybeam. We obtained a 5-$\sigma$ detection limit for the column density, for a line width of 16.5 \kms, of $3.3 \times 10^{18}$ cm$^{-2}$ for the lowest-resolution data cube and $1.2 \times 10^{19}$ cm$^{-2}$ for the highest-resolution data cube. The detection limit for small \HI\ clouds is about $6.5\times10^6$ $M_\odot$ (5$\sigma$ for a width of 16.5 \kms). Total intensity  \HI\ maps were made  by using masks based on a 2-$\sigma$ cutoff of data cubes where the resolution was degraded by a factor 2.

\begin{figure}
\centering
\includegraphics[width=8cm]{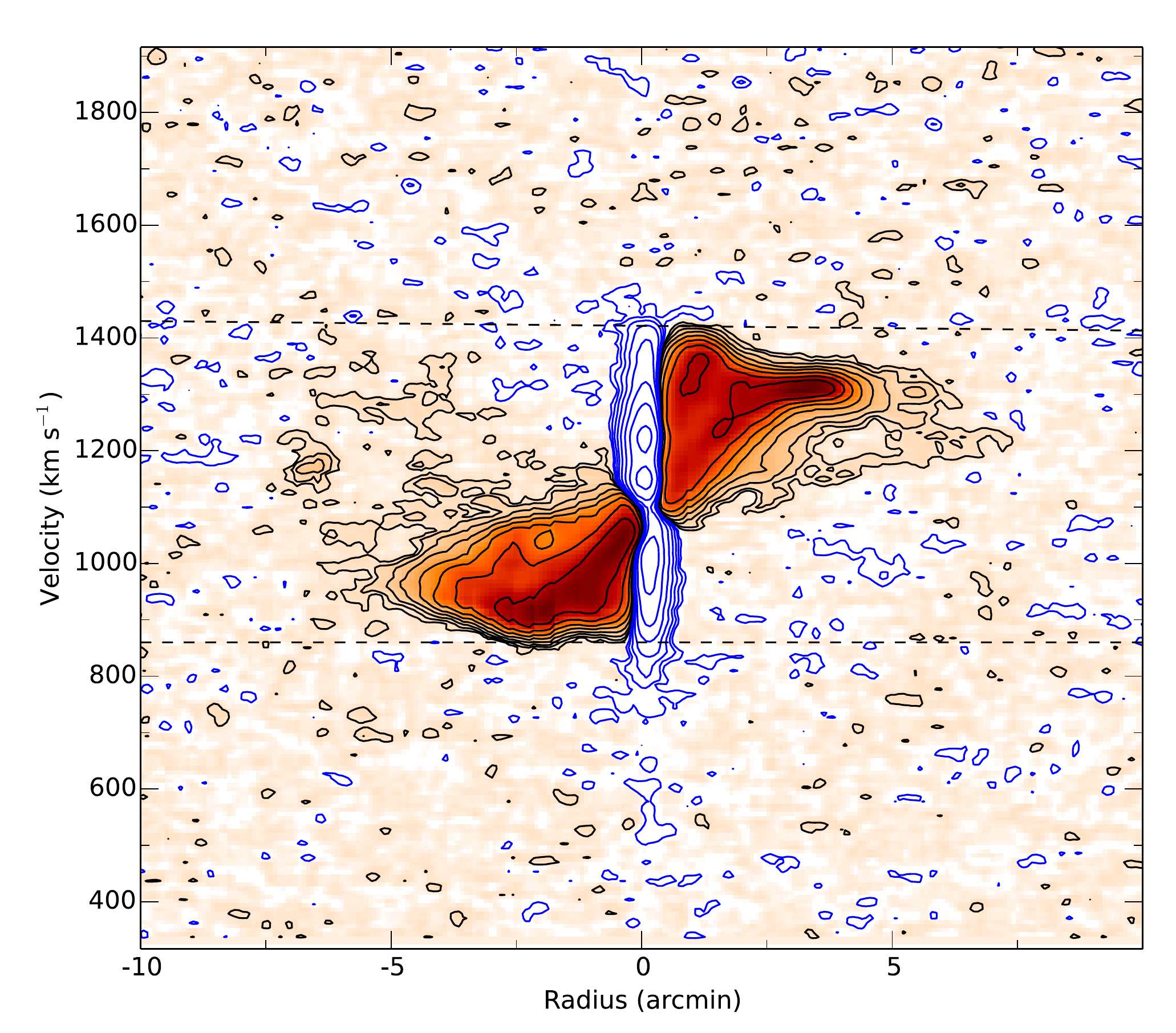}
\caption{Position-velocity slice taken along the major axis (P.A.=$166^{\circ}$). Contours are 1, 2, 4, 8, 16, 32, 64, 128 and 256 times 0.225 \mJybeam, and similarly for the negative contour levels (also shown as solid contours).  The NW side of the galaxy approaching. The dashed lines correspond to the range of velocities covered by the \HI\ in emission and are marked for reference in Fig. \ref{f:abs}  to be compared with the width of the broad \HI\ absorption. }
\label{f:abspv}
\end{figure}

\section{Results}\label{s:result}

\subsection{The H{\,\small I}  morphology and kinematics}

Figure \ref{f:totHI2} illustrates the total \HI\ intensity map showing the distribution of the gas in and around NGC~3079. The \HI\ column density contours are overlaid onto the Digitized Sky Survey (DSS2) optical image. Companions detected in \HI\ are also marked. For  a summary of the basic \HI\ parameters see Table \ref{t:parameter}. 
Figure \ref{f:abspv} shows the position-velocity (\textit{p-v}) map taken along the major-axis of NGC 3079. The map shows a clear pattern of rotation, with the northern side approaching and south receding.  Very striking is the very asymmetric kinematics at all scales. This is further discussed in Sec.\ \ref{s:disc}.

\begin{table*}
 \caption{\HI\ parameters of the NGC~3079 Group.}\label{t:parameter}
\begin{center}
\begin{threeparttable}
\begin{tabular}{c c l c  c  c }
\hline\hline
$\alpha$(J2000)&$\delta$(J2000) & Name& $V_{\rm sys}$ &$M_{\rm HI}$ & Label \\
(h m s)&($^{\circ}$ $^{\prime}$ $^{\prime\prime}$)& (NED) &(\kms)&$(10^{8}$ \msun)& \\
\hline
10 01 57.8 & +55 40 47 & NGC~3079                  & 1116 & 96.3 & - \\
10 00 48.2 & +55 36 56 & NGC~3073                  & 1171 & 1.86 & - \\
10 01 16.8 & +55 42 58 & MCG~9-17-19               & 1289 & 3.42 & - \\
10 03 10.0 & +55 35 56 & J100311.18+553557.6\tnote{n} & 1050 & 0.09 & A \\
10 03 32.5 & +55 31 19 & J100331.69+553121.1\tnote{r}& 1365 & 0.17 & B \\
10 01 20.7 & +55 54 21 & [YGK81] 078 \tnote{n}  & 1164 & 0.16 & C \\
09 59 42.9 & +55 53 16 & J095940.91+555317.8\tnote{n} & 1246 & 0.14 & D \\
10 04 43.8 & +55 19 46 & J100444.00+551943.2\tnote{r} & 1148 & 0.57 & E \\
\hline
\end{tabular}
\hspace*{0.2cm}Note: Objects are labeled according to Fig.\ \ref{f:totHI2}
	\begin{tablenotes}
	  \item[n]SDSS galaxy with a maximum separation of $0.2^{\prime}$ but no redshift information available.
	   \item[r]SDSS galaxy with a matching redshift.
\end{tablenotes}
\end{threeparttable}
\end{center}
\end{table*}

The improved sensitivity of our observations has allowed to detect new, faint and kinematically distinct \HI\ features compared to the previous observations.  The \HI\  total intensity map of NGC~3079 shows the \HI\ disk to be far more extended and disturbed than previously observed \citep{b32}. The \HI\ is seen out to a projected radius of $\sim $22 kpc. The integrated \HI\ flux is $\sim$$161$ Jy \kms, implying a total \HI\  mass of $\sim$$9.6\times 10^{9}$ \msun. The improved sensitivity of the new data highlights the more extended \HI\ distribution resulting in $\sim$25\% more \HI\ detected compared to what reported by \citet{b32}. This difference is particularly due to the much larger extent of the galaxy west of NGC 3079  in the new data. Several new \HI\ extensions and plumes of emission are also visible to the south-west and north-east.  It is interesting to note that in the previous data the galaxy appeared more extended towards the east than to the west, while our deeper observations now show that the reverse is true and the \HI\ is much more extended to the west, in the direction of MCG 9-17-9.

Figure \ref{f:chan} shows selected velocity channel maps (every second channel) containing \HI\ emission detected in NGC~3079. Several key properties of the \HI\  distribution and kinematics can be inferred from these images. The emission spans over a wide velocity range from $\sim$850 to $1410$ \kms, with a systemic velocity of $\sim$1116 \kms.  The channel maps show that most of the extended \HI\  is located in the north-west part of the galaxy. Furthermore, the \HI\ disk appears  warped on both north and south sides. This can be seen at velocities, e.g., $v=952$ and $v= 1281$ \kms. 
At lower levels ($2\sigma$) there is a sign of a trailing stream of gas visible at  velocities near   $1050$  \kms. This tail-like feature is likely produced by an interaction with the companions seen towards south-east of NGC~3079  (see Fig. \ref{f:totHI2}). Other streams and counter-rotating \HI\  are also visible velocities between $1116$ and $1165$ \kms. From these channel maps it is apparent that NGC~3079 is disturbed and has a lopsided nature allover the disk. 

There also seems to be an \HI\ stream associated with the galaxy seen north of NGC~3079 (([YGK81] 078; labeled C in Fig.\ \ref{f:totHI2}); visible for velocities $v=1149$ and $v= 1182$ \kms). Although the outer contours in the channel maps look disturbed, the extended gas  still shares the disk kinematics (e.g., $v= 1182$ \kms). The more likely source of the disturbance is an interaction between the two galaxies (see Fig.\ \ref{f:totHI2}). 
Perhaps, the most interesting feature in our data is the new detection of large stream of gas encircling the galaxy (Figs.\ \ref{f:totHI2} and \ref{f:chan}). This will be discussed in Sec.\ \ref{s:group} in relation to the companion galaxies .

\subsection{The H{\,\small I} absorption}\label{sec:absorption}

As expected from previous studies  \citep[e.g.,][]{b14}, \HI\ absorption is detected against the  nuclear continuum emission. The position-velocity diagram presented in Fig.\ \ref{f:abspv} shows the    absorption as blue contours. The deep part of the absorption profile  exhibits a multiple-component structure with a mixture of broad and narrow components, consistent with what found by previous studies at higher resolution \citep{b14,b21,swada00}.  The strongest absorption  extends in velocity from $\sim$850 \kms\ to $\sim$1440 \kms. 
The peak optical depth is high, $\tau \sim 0.18$, even when  assuming that the absorption is observed against the entire continuum  source in the inner region (i.e.\ the covering factor $f = 1$). This corresponds to a column density of $N_{\rm HI} \sim 2.0 \times 10^{22}$ cm $^{-2}$ for a spin temperature of $T_{\rm spin}$ = 100 K. 

\citet{b14} have studied this absorption using high-resolution observations. Their detected \HI\ absorption covers a similar velocity range and shows an even higher optical depth,  suggesting that the covering factor for our data is not unity and that the absorbing screen is concentrated in the nuclear regions (i.e.\  not in front of the diffuse continuum emission  which is included in the relatively large central beam of our observations). 
\citet{b14} identified three distinct absorption components associated with the nuclear disk, the inner and the outer disk structures. 

\begin{figure}
\centering
\includegraphics[width=8cm]{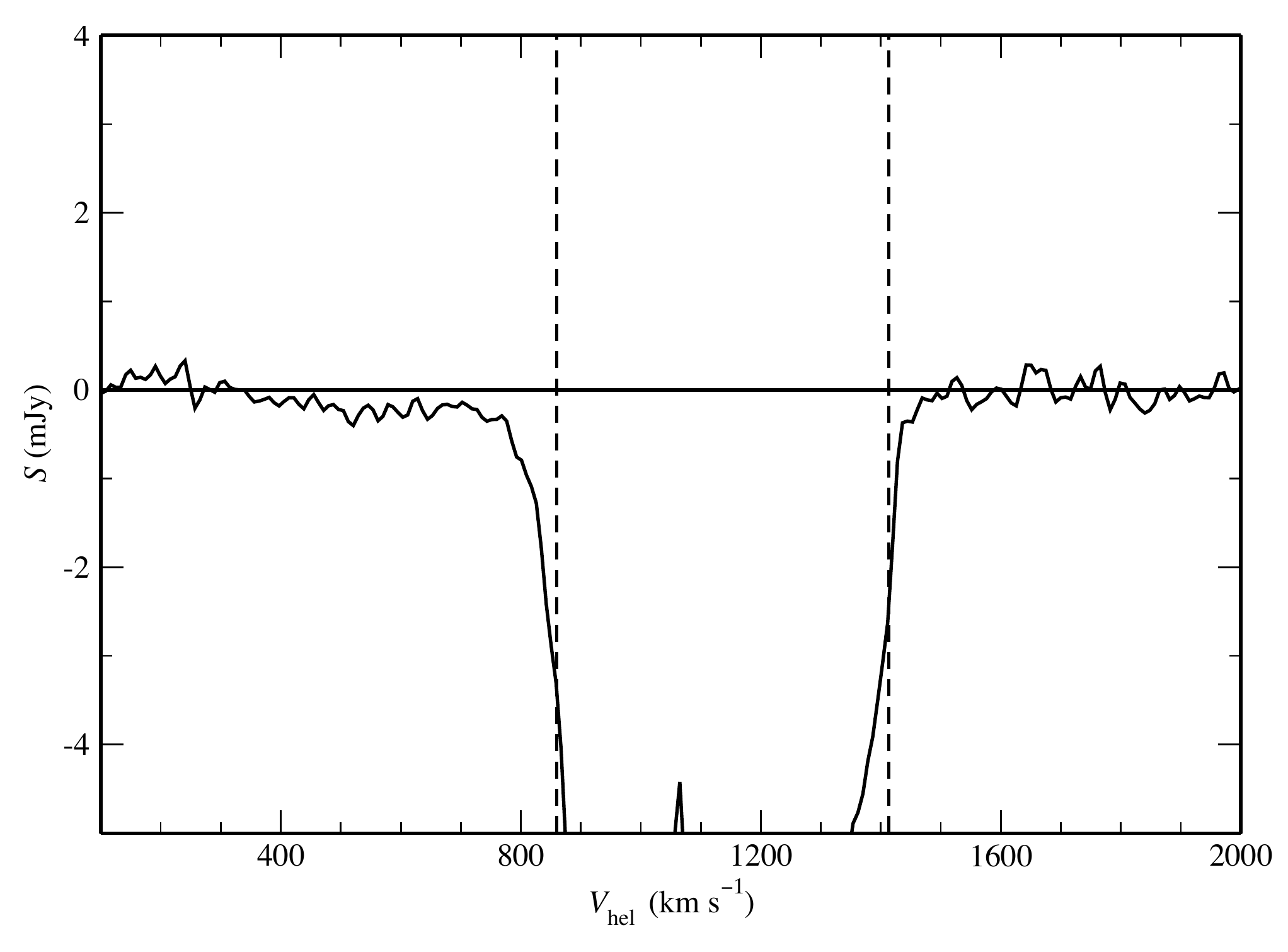}
\caption{A zoom-in plot of the \HI\ absorption. The vertical line represents the systemic velocity. The dashed lines represent the velocity range of the \HI\ emission as indicated in Fig.\ \ref{f:abspv}.}
\label{f:abs}
\end{figure}

Although our observations cannot provide a comparable spatial resolution, the sensitivity and bandwidth  of our data have allowed us to explore the presence of \HI\ (in absorption) not part of a regularly rotating structure. 
In addition to the deep absorption covering the range of rotation velocities of the \HI\ emission, we  detect a faint, additional component  that covers a velocity range between $\sim$450 \kms\ and  800 \kms,  well outside the velocity range of the regularly rotating disk. At the extreme velocities,  it is blueshifted by $\sim$600 \kms relative to the systemic velocity.  
To give a better view of this faint absorption component, a zoom-in of the \HI\ profile at the location of the peak continuum (which has a flux density of 310 \mJybeam) is shown in Fig.\ \ref{f:abs}. Interestingly, evidence for a  blueshifted \HI\  component was also reported by \citet{Gallimore1994} based on their VLA observations. The kinematics of the gas producing this shallow component is not consistent with gas in a simple rotating disk, rather it indicates the presence of outflowing gas. 
Since both starburst winds and AGN are known to be able to drive outflows, the superwind or the sub-arcsec radio jet might power such this outflow in this galaxy. Both mechanisms are discussed in Sec.\  \ref{s:outflow}.

This broad and shallow component   has a maximum depth of $\sim$0.35 mJy. Broad and shallow absorption components can be the result of errors in calibration caused by, for example, temporal instability of the bandpass of the telescope. In order to exclude this possibility, the stability of the bandpass was carefully verified by investigating whether similar broad features  appear in the spectra of the calibrators when the bandpass derived from one calibrator observed before  NGC 3079 was applied to a calibrator observation taken after the science observation. In this way we find that the level of instrumental variation of the bandpass is a factor ten below what would be needed to explain the faint absorption component, which implies that the broad absorption is not an instrumental effect. 

We can estimate the parameters of the broad absorption. If we assume that this component is observed against the peak of the continuum, its optical depth is $\tau \sim 0.0011$. Because there is a large uncertainty when using higher resolution data due to the variability of the core radio flux density reported in some observations \citep[e.g.,][]{swada00,kondra05,middel07}, we only use this value of $\tau$ to derive a  lower limit to the column density. The column density is estimated as $N_{\rm HI} = 1.82 \times 10^{18}\, T_{\rm spin} \int \tau\, d v$ cm $^{-2}$. This gives us  as lower limit of  $N_{\rm HI} \sim 4 \times 10^{19}$ cm $^{-2}$ (for $T_{\rm spin}$ = 100 K and covering factor $f = 1$). It might well be that, if the absorbing gas is located close to the centre of NGC 3079, that the spin temperature of the gas is well above the 100 K assumed here and that these column densities have to be increased accordingly. 

\subsection{The radio continuum}\label{s:cont}

\begin{figure}
\includegraphics[width=8cm]{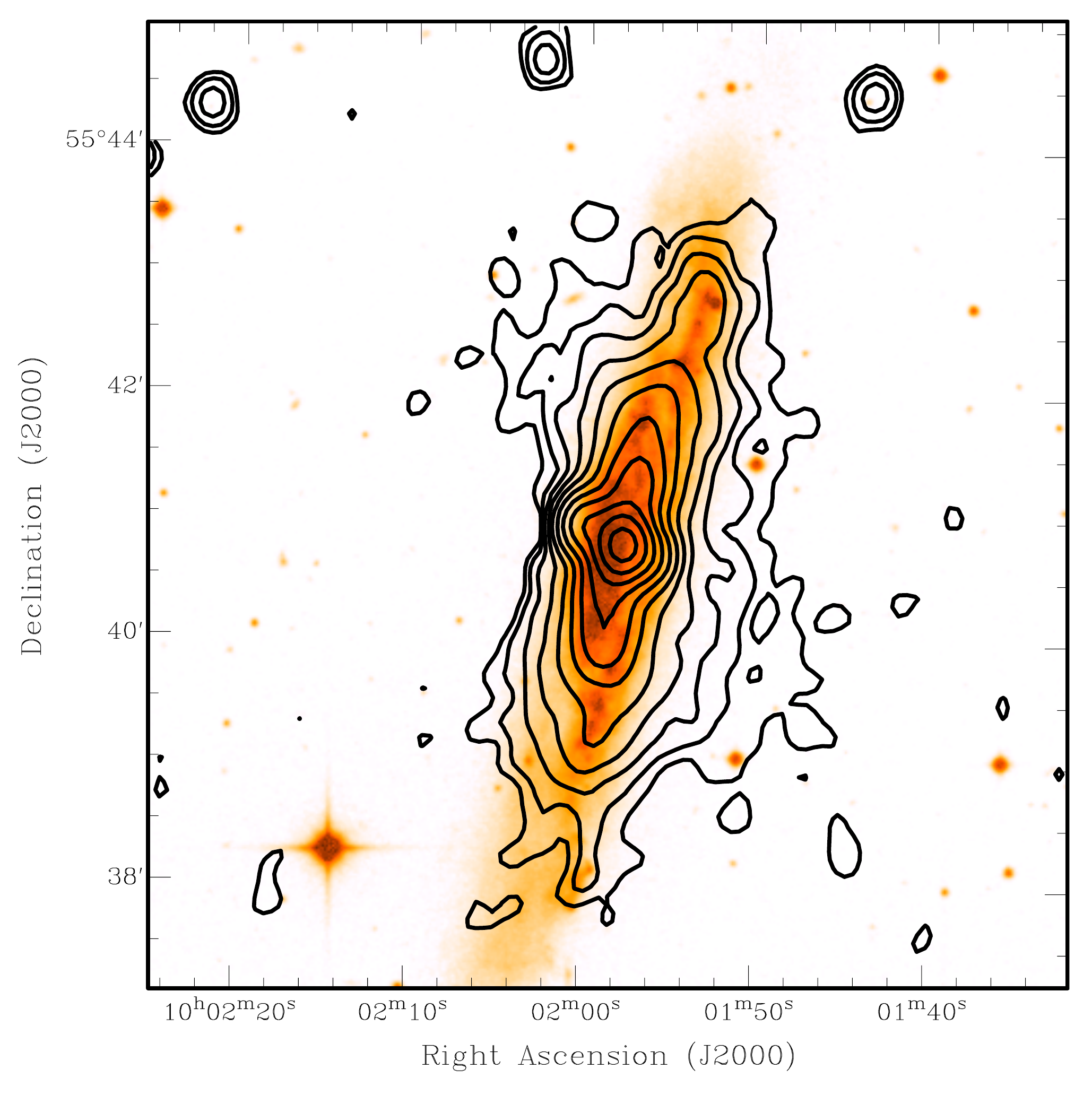}\label{f:after}
\caption{Continuum image of NGC~3079  (after ``peeling") overlaid onto an DSS2 image. Lowest contour level is 0.2  \mJybeam\ with following contour levels increasing  factors of 2.}
\label{f:cont}
\end{figure}

Figure \ref{f:cont}  shows the   continuum image of NGC~3079 overlaid onto the optical DSS2 red image.  The continuum structure of NGC~3079 has been well studied at various resolutions and frequencies \citep{b18,b14,irwin03}. At high spatial resolution, it shows two radio lobes (extending about 2 kpc from the nucleus) extending perpendicularly to the plane of the galaxy. These relatively large, extended radio lobes, likely connected to the nuclear outflow, are quite unusual for Seyfert galaxies. 
On larger scales, \citet{irwin03} found that the radio continuum follows the galaxy disk. This emission is surrounded by a radio  halo (extending at least 4.8 kpc from the disk) together with other extension/loops particularly clear at low frequencies .  

At the resolution of our observation ($\sim $1 kpc), we see some indication for  the nuclear lobes in the form of small extension perpendicular to the disk, although these cannot be well separated from the extended disk emission. The peak flux density in the centre of NGC 3079 is 310 mJy beam$^{-1}$ (see Fig.\ \ref{f:cont}). 

The disk emission  extends to radii similar to what observed by \citet{irwin03}. One outcome from our radio continuum data is that the halo appears asymmetric, with the emission on the W side being more extended than on the E. 
On the E side, the extent is about 5 kpc while on the S-W side it reaches about 7 kpc.

This extension to the S-W side was also seen in the 1.2~GHz image from \citet{irwin03}, while the remaining  emission looks smoother (and larger) in our image. This is possibly the result of the improved quality of our image thanks to our direction dependent calibration. We do not see evidence for the radio loops seen at low frequencies by \citet{irwin03}. The larger and asymmetric extent of the continuum halo on the W side partly mimics that of the the \HI\ emission. 
No continuum emission is detected, down to the level of 0.24 \mJybeam\ ($\sim$$3 \sigma$), from  companion galaxies.

\begin{figure}
\centering
\includegraphics[width=8cm,height=6cm]{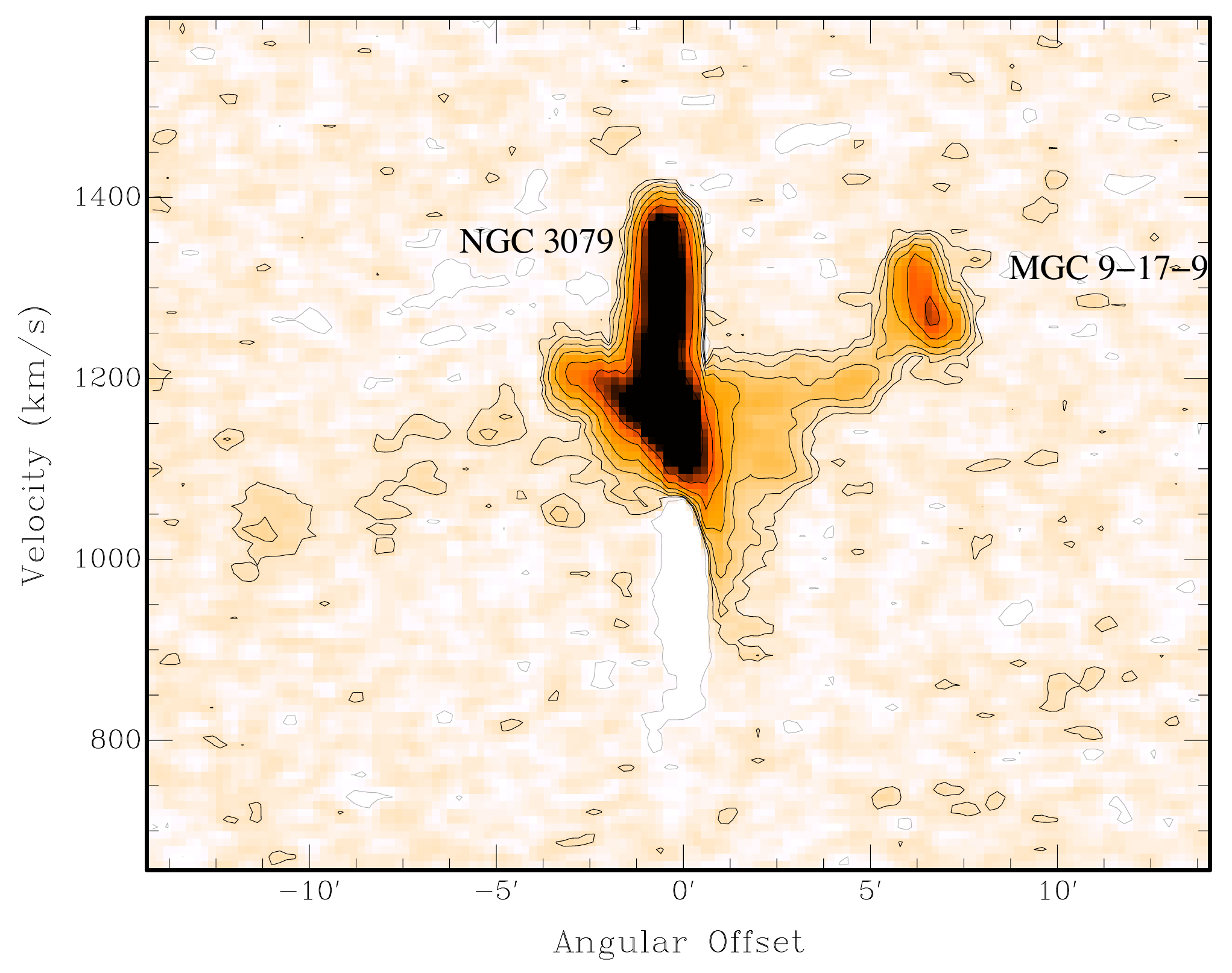}
\caption{Position-velocity slice taken along a line through the centres of  NGC~3079 and  MCG~9-17-9 (P.A.=$290^{\circ}$) showing the bridge between these two galaxies. The figure also shows the \HI\ extension towards J100311.18+553557.6. Contour levels are --0.3, 0.3, 0.6, 1.2, 2.4, ... \mJybeam. }
\label{f:stream}
\end{figure}

\subsection{The NGC~3079 Group}\label{s:group}

The NGC~3079 group is known to contain a number of smaller galaxies. Our data revealed \HI\ gas associated with seven galaxies that are close (spatially and in velocity) to NGC~3079, and thus are  likely  associated with the group (see Fig.\ \ref{f:totHI2} where the objects are labeled). Among these, four have previously known redshifts. All newly identified members of the group are small, gas-rich galaxies.
The  \HI\ parameters of  each galaxy are  given in Table \ref{t:parameter}.  
A velocity dispersion of  $\sigma \sim 80$ \kms\  is measured for the group. Four of the group members (MCG~9-17-9, NGC~3073, SDSS~J100311.18+553557.6 (labeled A) and [YGK81] 078 (labeled C)) are seen interacting with NGC~3079.  
Below, we briefly discuss some details of these interactions.

\subsubsection{MCG~9-17-9}\label{s:mcg}

MCG~9-17-9 is a small spiral galaxy located about 6.5 arcmin ($\sim$33 kpc)  north-west of NGC~3079 (see Fig.\  \ref{f:totHI2}). It is classified as Sb-Sc, and seen fairly face on. Very little is known about this galaxy other than from observation of the NGC~3079 environment \citep[e.g.,][]{b1,piets98}.  Bright X-ray emission ($L_{\rm X} \sim 1.59 \times 10^{39}$ erg~s$^{-1}$) has been detected by \citet{piets98} which was considered rather high for such a small spiral galaxy. The \HI\ appeared undisturbed in the interaction of the group in \citet{b1}. The new \HI\ data, however, reveal that an interaction between MCG~9-17-9 and NGC~3079 is ongoing. This is evident by the \HI\ `bridge' detected between the two galaxies that overlaps both spatially (in projection) and in velocity space, suggesting a physical link between the two galaxies.
This can be seen  clearly in Fig.\ \ref{f:stream} where the \textit{p-v} slice along the bridge is shown. The south end of the bridge joins the north-west part of the \HI\ disk of NGC 3079 which also seems disturbed in that region.  The other extreme of the bridge connects to MCG 9-17-9.  Interestingly, no perturbations are seen even in very deep optical images (P.-A. Duc, priv. comm.), suggesting the bridge is not tidal in nature.  We discuss this in more detail below.

The \HI\ bridge appears to extend to the other  side of NGC~3079, towards the other small galaxies SDSS~J100311.18+553557.6, and SDSS~J100331.69+553121.1 (marked A and B respectively in Fig.\ \ref{f:totHI2}). While it is possible that this means that the bridge extends this far, it is also possible that this \HI\ forms another tail, emanating from SDSS~J100311.18+553557.6 because this galaxy seems to have an \HI\ tail pointing towards NGC~3079.  
Overall, the large streams of gas encircling NGC~3079  resemble some of the \HI\ streams found in group galaxies, such as Magellanic Stream \citep[e.g.,][]{Wannier1972}.

\subsubsection{NGC~3073}
\begin{figure}
\centering
\includegraphics[width=7cm,angle=-90]{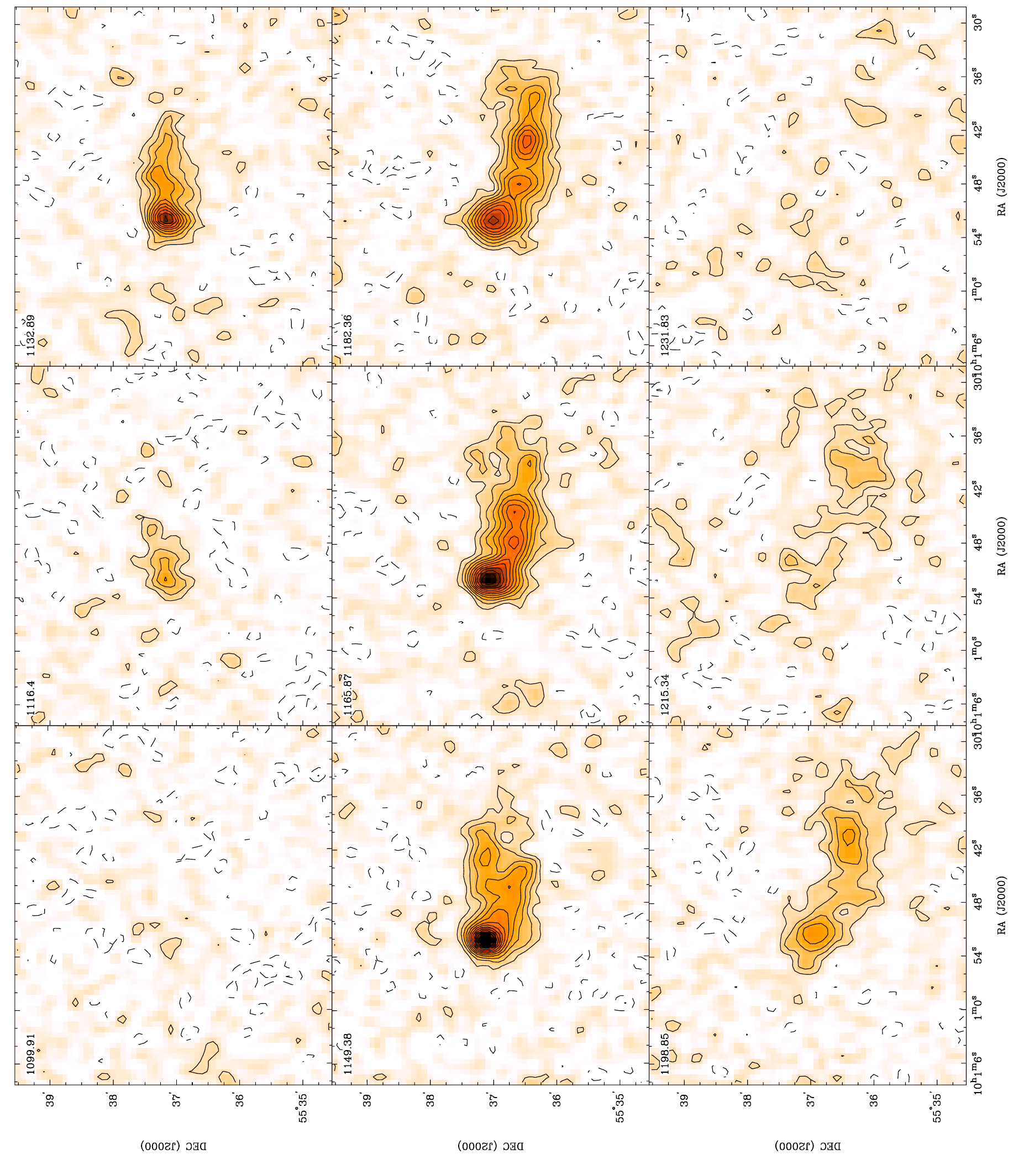}
\caption{Channels maps from the highest resolution cube ($19^{\prime\prime}\times23^{\prime\prime}$) showing the two \HI\ tails of NGC 3073.  The lowest contour level is 0.25 \mJybeam\ which is also the increment of the contour levels. The dashed contour corresponds to --0.25 \mJybeam.}
 \label{f:2tails}
\end{figure}

\begin{figure}
\centering
\includegraphics[width=8cm]{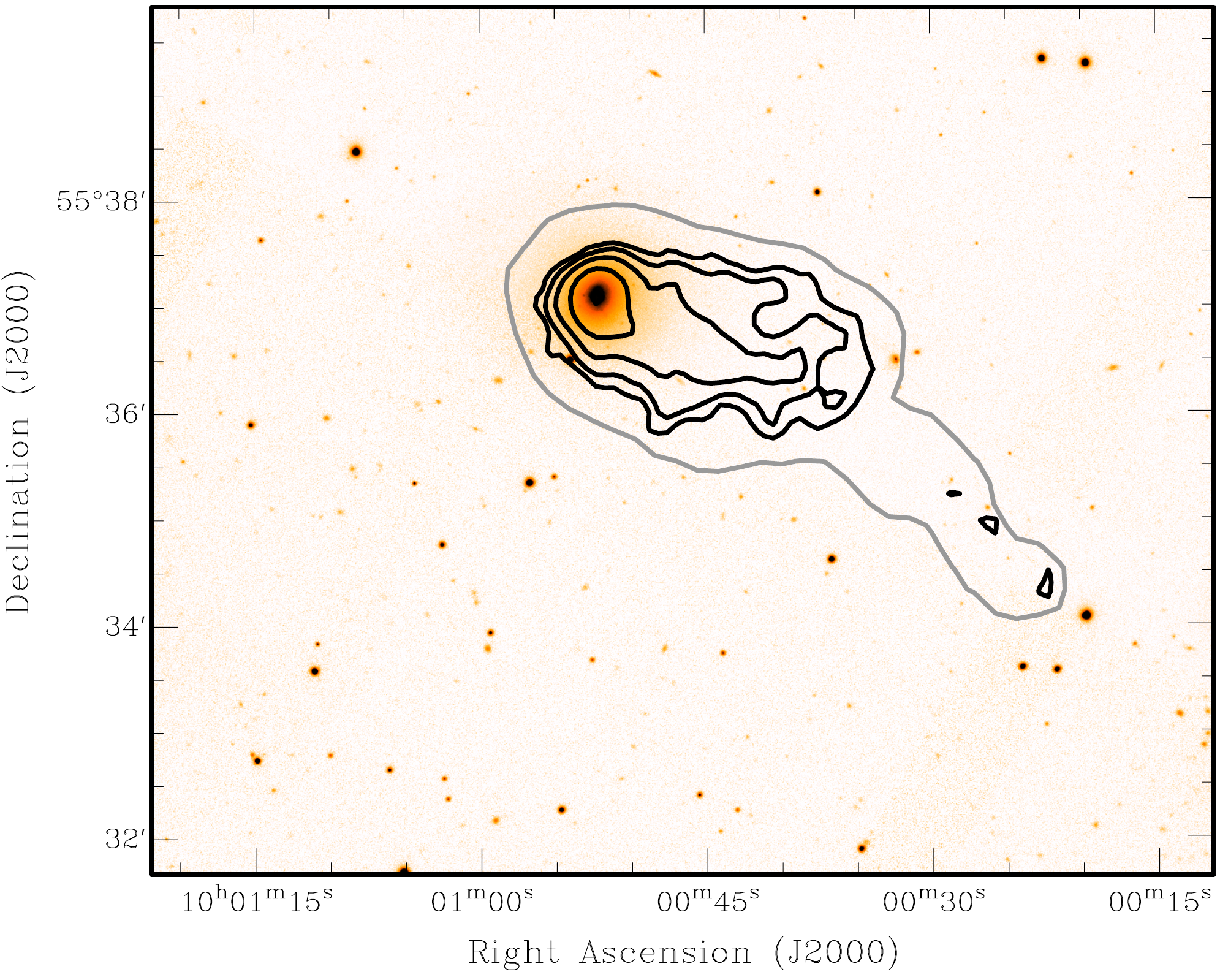}

\caption{Integrated \HI\ image of NGC 3073 showing the full extent of the two tails of NGC~3073. Black contour levels are at 3, 5, 10, 20, 50, 100 $\times 10^{19} \mathrm{cm}^{-2}$ ($45^{\prime\prime} \times 48^{\prime\prime}$ resolution), and the grey contour level is $1.5 \times 10^{19} \mathrm{cm}^{-2}$ ($78^{\prime\prime} \times 80^{\prime\prime}$ resolution).}
 \label{f:tail}
\end{figure}

NGC~3073 is a dwarf SAB0 galaxy located about 10 arcmin ($\sim$ 50 kpc) to the west of NGC~3079. The optical spectrum of NGC~3073  shows a young stellar population and high levels of star formation for its kind  \citep{Ho2009,McDermid2015}. \citet{b1} revealed an elongated \HI\ tail in this galaxy pointing away from NGC 3079, but aligned with the nucleus of NGC~3079. The total \HI\ map of this galaxy from our data is shown in Fig.\ \ref{f:tail}. The \HI\ tail is clearly visible and turns out to be even longer than in \citet{b1}. However, the tail  also seems to consist of two streams. This can be seen in the channel maps shown in Fig.\ \ref{f:2tails} (from 1132 to 1198 \kms) and in Fig.\  \ref{f:tail}. The significant curvature of the tail is also visible in Fig.\ \ref{f:tail}. The tail extends to a projected length of about 5 arcmin (25.5 kpc) and it bends  at the end giving a ``cometary" appearance to the galaxy. 

Although we discuss the \HI\ tail of NGC 3073 in more detail below (Sec.\ \ref{s:discn3073}), the presence of the secondary branch may indicate that ram pressure stripping is the cause of the tail. The two-tailed structure of the \HI\ tail of NGC 3073 shows a fair degree of similarity with that of NGC 4330 \citep{chung07} for which it is well established that it is caused by ram pressure stripping by the hot medium of the Virgo cluster.  

\section{The structure of the H{\,\small I} disk }\label{s:disc}

\begin{figure}
\centering
\includegraphics[width=8cm]{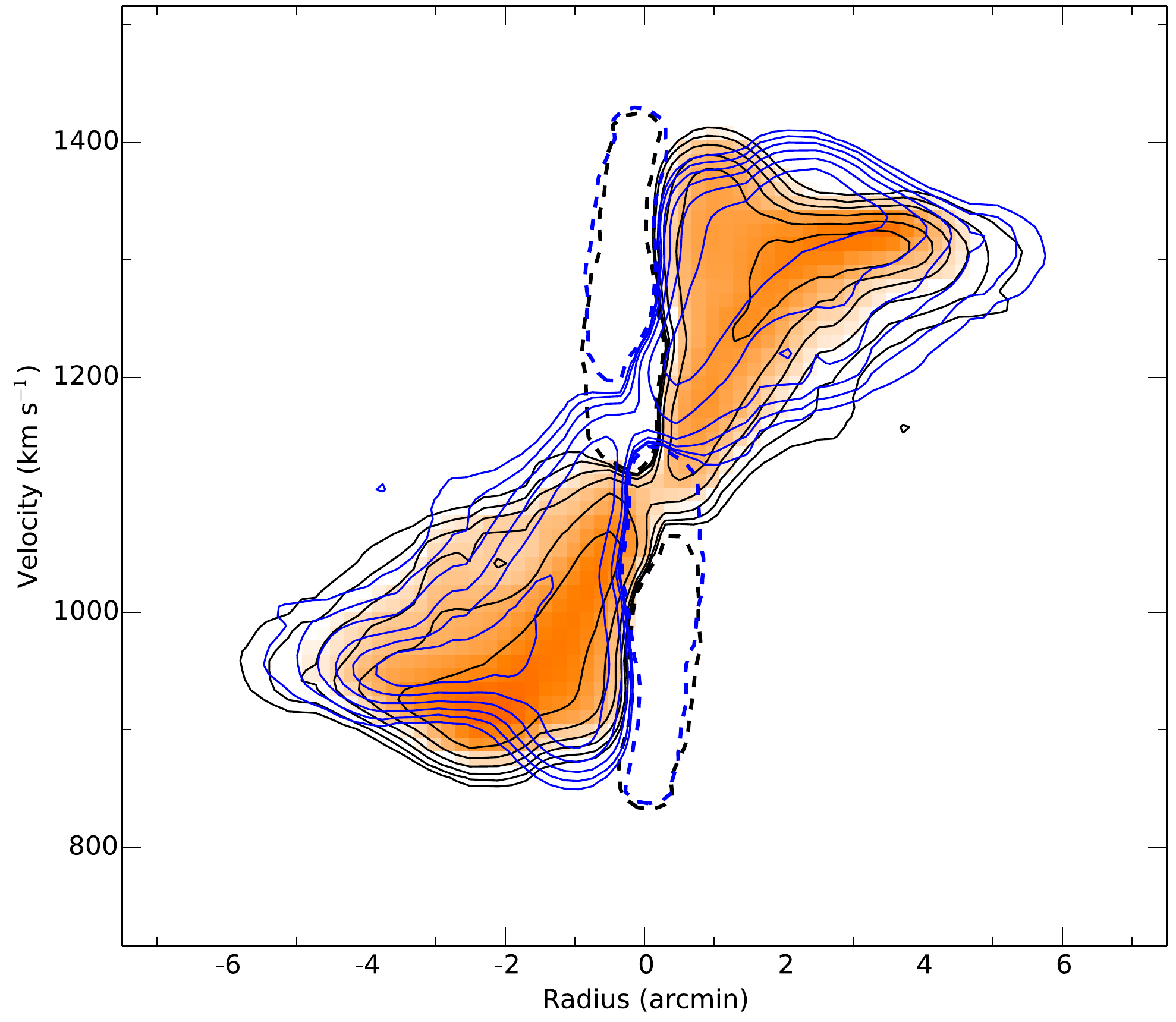}
\caption{Plot illustrating the lopsidedness of NGC~3079 (through position-velocity slices taken along the major axis). One side of the galaxy has been flipped, both spatially and in velocity, on top of the other. It is clear that the kinematics of the disk is very asymmetric at all radii.}
\label{f:flipped}
\end{figure}

Our deep \HI\ observations of NGC~3079 and its group  reveal several new features, in and around the galaxy, that are clear signatures of the complex and hectic history of this object.  The presence of a number of tails and streams, as well as the warped and lopsided structure and kinematics of NGC 3079,  point to the strong effects of  interactions with the   environment. Although the kinematics of the \HI\ disk of NGC 3079 is dominated by regular rotation, it appears to have gone through a major ``shake up" evidenced by its lopsidedness and warped structure. 
Apart from the morphological lopsidedness, in particular visible in the outer regions, Fig.\  \ref{f:abspv} shows that NGC 3079 also has a strong kinematical lopsidedness in the bright disk. This  is emphasised in Fig.\ \ref{f:flipped} where the kinematics of the two sides of NGC 3079 are compared. It is clear that the kinematics of the disk is very asymmetric at all radii.
The lopsidedness is seen over a large range in radius, but is particularly strong in the central regions. This non-standard kinematics, with associated non-circular motions, may well be related to the elevated activity in the central regions (elevated star formation and AGN). 

The distribution of the \HI\ is asymmetric in the north-south as well as in the east-west directions, i.e.\ more extended towards the north and west compared to south and east sides of the disk. This gives the galaxy a lopsided appearance. As can be seen in  Figs.\ \ref{f:totHI2} and \ref{f:flipped}, the morphological lopsidedness  of the disk appears to begin near the edge of the optical disk where the \HI\ surface density starts declining, and slowly increases at larger radii. In particular in the direction of MCG 9-17-9 the disk has a large extension.

Figure \ref{f:abspv} shows the position-velocity (\textit{p-v}) diagram taken along the major axis. The disk shows a clear pattern of rotation, with the northern side approaching and south receding.  Apart from this overall pattern of rotation, several streams of gas can be seen in this figure. In particular in the northern (approaching) side of the galaxy, gas is found over a wide range of velocities outside the range of the rotating disk. Similarly, on the receding side gas is found at anomalous velocities. This likely corresponds to a stream of gas  at large  radii.

Lopsidedness in the mass and gas distributions is a well known phenomenon and affects a significant fraction of galaxies \citep[e.g.,][]{Richter1994,Haynes1998,Sancisi2008} and is generally interpreted as a sign of interaction and/or accretion. Similarly, warping is also common among disk galaxies and is known to affect many galaxies where the gas disk extends well beyond the stellar distribution \citep{Ruiz2002,Jozsa2009}. The warp in the outer disk (north and south) appears fairly mild but detailed modelling will be needed to determine the relative importance of the warp. 

Considering the complicated structure of the gas in and around the galaxy, the striking bridge between NGC 3079 and MCG~9-17-9, as well as the other gas streams in the direction of the newly discovered  companions ([YGK81] 078 and SDSS~J100311.18+553557.6), it is likely that the lopsidedness and the warp are induced by  interactions with companion galaxies via tidal encounters.

\section{The interplay between NGC~3079 and its rich environment}
\label{s:environment}

In this section, we discuss in some more detail the \HI\ tails connected to the companions galaxies MCG 9-17-9 and to NGC 3073.

\subsection{MCG~9-17-9: Tidally or Ram-Pressure Stripped?}\label{s:discmcg9}

The  interaction between MCG~9-17-9 and NGC~3079 is traced by an \HI\ bridge as illustrated by  Fig.\  \ref{f:stream}.  The total \HI\ image of MCG 9-17-9 shows that  the \HI\ bridge forms near this galaxy at a column density of about $2\times 10^{20}$ cm$^{-2}$ and a radius of about 3.3 kpc. However, Fig.\  \ref{f:stream} also shows that the bridge connects to the distorted outer disk of NGC 3079, suggesting that some part of the gas of the entire complex originates from the disk of NGC 3079. The two galaxies have a relative velocity of 173 \kms\ with MCG~9-17-9 being red-shifted with respect to NGC~3079. A velocity change of $\sim$100 \kms\ is seen along the bridge. From the size of the bridge and the relative velocities, one finds a kinematical timescale ($R/V$) of  $\sim$$ 1.9 \times 10^{8}$yr, giving an indication of the age  of the bridge. This timescale is not very different from the  age of the starburst in NGC~3079  estimated by \citet{veille94} (i.e.\ $10^{7}$-$10^{8}$yr), suggesting that the interaction between MCG~9-17-9 and NGC 3079 may have played a role in triggering the starburst. Here, we discuss the possible origin of the \HI\ bridge.

\subsubsection{The case for tidal stripping}

\HI\ bridges or streams between galaxies are often associated with tidal encounters. The fact that the outer disk of NGC 3079 seems to be affected by the passage of MCG 9-17-9 does indicate that tidal effect, at some level, play a role. However,  very deep optical images (P.-A. Duc, priv. comm.) do not show any indication for a tidal distortion of the stellar distribution of MGC 9-17-9. This makes tidal effects unlikely to be  the main cause of  the \HI\ bridge. Moreover, a tidal encounter often leads to two arms, one leading and one trailing, while here we  observe only one  arm. That tidal effects are not the dominant origin of the \HI\ bridge is reinforced from an estimate of the tidal radius of MGC 9-17-9. The radius $r_{\rm t}$ beyond which matter is tidally stripped due to an interaction between two galaxies can be estimated using
\begin{equation}
r_{\rm t} > R\left( \frac{m}{2 M}\right)^{1/3}
\end{equation}
where $m$ is the mass of the satellite, $M$  the mass of the main galaxy and $R$  the projected separation between the two  (\citealt{binney87}). The un-equality is due to the fact that the projected separation gives a lower limit to the true separation.

The ratio  of the masses of the two galaxies  we estimate from their rotation velocities  and their relative sizes. For the rotation velocity at the edge of the regular \HI\ disk of NGC 3079 ($\sim$5 arcmin = 23.3 kpc radius) we find 200 \kms\ (see Fig.\ \ref{f:abspv}) while for MCG 9-17-9 we find 40 \kms\ at a radius of $\sim$0.7 arcmin = 3.3 kpc. This gives a mass ratio of 166. The projected separation between the two galaxies is 33 kpc. 

Using these values in Eq.\ 1 gives a tidal radius of $r_{\rm t} > 4.8$ kpc,  larger than the radius of the regular, unaffected disk of MCG 9-17-9. However, this estimate is not very much larger than the size of the disk of MCG 9-17-9 and it might well be that,   e.g.\ during those phases of the encounter between where the satellite may have been closer to NGC 3079, tidal effects have played some role. The distortion of the outer disk of NGC 3079 also points to this. However, given that the bridge connects to MCG 9-17-9, the formation of the bridge seems to be still ongoing and, given the estimated tidal radius,  it  is unlikely that  tidal effects are the main reason for the  \HI\ stream near MCG 9-17-9.

\subsubsection{The case for ram pressure}

Another scenario  is that the \HI\ bridge is caused by ram pressure stripping of \HI\ from MCG 9-17-9 by the halo of hot gas  NGC 3079 is known to posses \citep{strick204}.  

The gas in a galactic disk is stripped if the the ram pressure of the hot halo gas is greater than the restoring gravitational force per unit area of the galaxy's disk \citep{gunn72}. The density of the hot gas, $\rho_{\rm h}$, required for  stripping gas at a disk radius $R$ can be estimated,   assuming a gas disk sitting in a uniform spherical stellar/dark matter system,  from 
\begin{equation}\label{eq:ramp}
\rho_{\rm h}  \geq  \Sigma_{\rm g} V^{2}_{\rm rot}R^{-1}V^{-2}
\end{equation}
where $\Sigma_{\rm g}$ is surface mass density of the gas disk, $V_{\rm rot}$  the rotational velocity, $V$  the velocity of the galaxy through the hot halo, and $R$ the radius of  the disk under consideration \citep{gunn72}.

From the total \HI\ image, we see that the disk of MCG 9-17-9 appears disturbed for column densities below $\Sigma_{\rm g}= 2\times10^{20}$ cm$^{-2}$ at a radius $R$ of 3.3 kpc. For the rotation velocity we use, as before, 40 \kms, while for the velocity $V$ of the galaxy w.r.t.\ the hot medium we assume 200 \kms, somewhat larger than the observed velocity difference between NGC 3079 and MCG 9-17-9. This gives us an estimate for the density of the hot gas required for ram pressure stripping to occur at the column density level and radius observed of $7.9\times 10^{-4}$ cm$^{-3}$.

This value should be compared with estimates of the density of the hot halo of NGC 3079. An extensive X-ray halo, with a luminosity of $L_{\rm X} \sim 6 - 8 \times 10^{39}$ erg~s$^{-1}$ and a temperature of  $\sim 4 \times 10^{6}$ K, has been detected around NGC~3079 (ROSAT; \citealt{piets98}, Chandra; \citealt{Cecil2002,strick204}). From the parameters derived for an exponential distribution of the hot gas (well-fit) by \citet{strick204}, we get, for the location of MCG 9-17-9, an observed density of  $n_{\rm e}= 7.3 \times 10^{-4}f^{-1/2}_{\rm v}$  cm$^{-3}$, where $f_{\rm v}$ is a volume filling factor. A very similar density was derived by \citet{piets98} for the hot halo of NGC 3079 assuming a  simple spherical distribution. These densities are quite close to that required for ram pressure to occurring at the column density and radius we observe the \HI\ disk of MCG 9-17-9 to be affected.

From the above analysis of both tidal and ram-pressure stripping, we conclude that the stripping that is currently occurring for MCG 9-17-9 is mainly due to the interaction with the halo of hot gas known to surround NGC 3079. The observed densities of the halo match quite well the estimated required densities for the stripping to occur at the observed column densities and radius. The estimated tidal radius is larger than the size of the disk of MCG 9-17-9. However, the difference is not very large and it is conceivable that, during an earlier phase of the encounter where possibly MCG 9-17-9 was closer to NGC 3079, tidal effects may have been important. The distorted outer \HI\ disk of NGC 3079 also suggest that this is the case.

\subsection{The two tails of NGC~3073}\label{s:discn3073}

The `cometary' appearance of the \HI\ in NGC~3073 was first observed and discussed by \cite{b1}. Our deeper data reveal that the \HI\ tail of NGC 3073 is not only more extended, but also that it appears to be composed of two tails superimposed (Fig.\ \ref{f:tail}). 

Irwin et al.\ discuss the interesting possibility that this tail of NGC 3073 is due to ram-pressure stripping by the fast star-formation driven wind coming from the central regions of NGC 3079. This possibility was also discussed by \citet{strick204}. Before we discuss this option in more detail, we first consider the alternatives of tidal effects and ram-pressure stripping by the hot halo of NGC 3079.

The first option can likely be excluded. In the previous section we showed that tidal effects are currently not very strong for MCG 9-17-9 and there are a number of reasons why for NGC 3073 this is even more the case. In the first place because the distance between NGC 3079 and NGC 3073 is likely to be larger, given the larger projected separation of 50 kpc. Moreover, NGC 3073 is more massive than MCG 9-17-9, reducing the tidal impact of NGC 3079. Using the results from \citet{Cappellari2013}, we estimate the rotation velocity of NGC 3073 to be 86 \kms, compared to 40 \kms\ for MCG 9-17-9. Taking this together, we find that the tidal radius for NGC 3073 is likely to be larger than about 17 kpc, which is much larger than the size of the galaxy ($\sim$1.2 kpc radius), ruling out a tidal origin for the \HI\ tail.

Similarly, we can also rule out that  the general halo of hot gas is stripping the \HI\ from NGC 3073. In an similar way as done for MCG 9-17-9, we estimate that the required density of the hot halo in order to be able to strip the gas from NGC 3073 is $1.3\times 10^{-2}$ cm$^{-2}$ which is much larger than the observed densities.

However, as \citet{b1} remark, it is possible that the ram pressure from the wind coming from NGC 3079 is stripping NGC 3073. This is mainly, since ram pressure depends on velocity as $v^2$,  because of the high velocity of the wind (1000 \kms\ \citealt{Cecil2001}). For this velocity, the required density to strip the \HI\ at the column density observed ($2.5\times10^{20}$ cm$^{-2}$), is about $7.5\times10^{-4}$ cm$^{-3}$.
The lower limit for the electron density of the super-wind, near the starburst in the centre of NGC 3079, is  estimated  to be $n_{\rm e}=2.7$ cm$^{-3}$ (Strickland et al.\ 2004). Assuming that the density of the wind declines with distance from the starburst according to $r^{-2}$, we obtain a density of about $1 \times 10^{-3} \mathrm{cm}^{-3}$ at the location of NGC 3073. Given the uncertainties, this matches the required density for stripping and we reiterate the suggestion made by \citet{b1} that the cometary \HI\ tail of NGC 3073 is due to ram-pressure stripping by the fast, starburst driven wind from the centre of NGC 3079.

\section{The interplay between the central activity and the H{\,\small I} disk}
 \label{s:outflow}
 
The finding of the \HI\ outflow seen in absorption can be added to the many interesting features of the nuclear region of NGC~3079. The broad and shallow blueshifted \HI\ absorption component  is blueshifted up to 600 \kms\ with respect to the systemic velocity. Being detected in absorption, this ensures that the gas is in front of the radio continuum and the blueshifted velocities indicate that the absorbing gas is outflowing. Such a broad, blueshifted component  has been tentatively seen also in  the VLA observations of \citet{Gallimore1994}. A component with such large velocities must be  associated to extreme, non-gravitational radial motions. 
 
Fast \HI\ outflows have been detected in several other radio sources with a range of AGN power. Many cases detected so far have been found to be associated to powerful radio jets (e.g., \citealt{Morganti2005}, 2013), although cases associated to low-power radio sources (e.g., NGC~1266; \citealt{Nyland2013}, Mrk~231; \citealt{Morganti2011}) are also known.
In a number of these objects, the most likely origin for the outflow is considered to be the kinetic push by the radio plasma jet following its interaction with a dense ISM \citep[e.g.][]{Morganti2013}. This has also been suggested to occur in lower-power radio sources such as NGC~1266 \citep{Nyland2013}, a source with a radio power more similar to that of  NGC~3079.    

In the case of NGC~3079, it is hard to identify the origin of  the outflow. This is due to the difficulty  to disentangle the effects due to the starburst from those caused by the AGN, as it has been discussed in detail for the outflow of ionised gas  by \citet{veille94} and \citet{Cecil2002}. According to these studies, the starburst component is the most likely driver, although an AGN component cannot be excluded.
Indeed, while the high infrared luminosity and the morphology of the well-known bubble observed in NGC 3079 suggest the stellar wind to be important\citep{veille94}, it has been suggested that the energy output of the AGN and starburst activities  are comparable \citep{Iyomoto2001}.

The possibility of the existence of a molecular outflow driven by the active nucleus was already suggested by 
\citet{haward95}.
Molecular outflows have now been observed in a number of objects. The multi-phase character of gas outflows, and the co-presence of molecular and atomic  hydrogen (as well as ionized gas), has been established in a growing number of cases where the outflows are driven by AGN. Thus, the \HI\ outflow detected in our observations may represent the counterpart of the outflowing molecular gas (e.g.\ as in the case of the radio-loud Seyfert 2 IC~5063, \citealt{morga15,Tadhunter2014}). 

The limited spatial resolution of our observations cannot add information about the precise location of the outflow. However, 
in order to shed light on its origin, we can at least derive the energetics of the outflow  and compare this with what is associated with the starburst and the non-thermal radio emission. This should tell us whether any of them can be ruled out because of  insufficient energy.

From the lower and upper limit of the column density as described in Sec.\ \ref{sec:absorption}, a range of values for the mass outflow rate can be obtained following  \cite{heck02} and \cite{rupk02}: 
\begin{equation}
\dot{M} \sim 30  \left ( \frac{\Omega}{4\pi}\right )  \left ( \frac{r_{*}}{1\ \mathrm{kpc}} \right )  \left ( \frac{N_{\rm HI}}{10^{21}\ \mathrm{cm}^{-2}}\right)\left( \frac{v}{300\ \mathrm{km}\ \mathrm{s}^{-1} }\right )M_{\odot}\  \mathrm{yr}^{-1} 
\end{equation}
where the gas is outflowing at a velocity $v$ from radius  $r_{*}$ over a solid angle $\Omega$. If we assume the outflow originates from the nuclear region, $r =100$ pc,  $\Omega$ to be $\pi$ steradians and $v= 500$ \kms, the above equation yields a lower limit to the mass outflow rate of $\dot{M}$ ranging from 0.05 to 0.5 $M_{\odot}$ yr$^{-1}$ where we have allowed $T_{\rm spin}$ to range between 100 K and 1000 K.
These mass outflow rates  are low compared those derived for ULIRGs (ranging between 10 to 1000 $M_{\odot}$ yr$^{-1}$, \citep{heck02,rupk02,Rupke2005a,Rupke2005b} and for radio galaxies, ranging between 1 and 50  $M_{\odot}$ yr$^{-1}$ \citep[e.g.][]{Morganti2005}. 

\citet{veille94} derived an upper limit for the ionised gas outflow in NGC~3079 of $\dot{M}<10 \times {n}_{\rm e}^{-1}$ \msun\ yr$^{-1}$  (for $10 < {n}_{\rm e} < 125$  cm$^{-3}$). 
Thus, the ionised gas outflow rate in NGC~3079 is relatively modest compared to ULIRGs, but of similar amplitude as the \HI\ outflow rates. 

With the \HI\ mass outflow rates, we can derive the kinetic power of the outflow using the approach of \cite{holt06}, including both the radial and turbulent component of the gas motion.
\begin{equation}
\dot{E}  \sim 6.34 \times 10^{35}\left(\frac{\dot{M}}{2}\right) \left(v^{2}+ \frac{\mathrm{FWHM}^{2}}{1.85}\right)    \mathrm{erg}\ \mathrm{s}^{-1}
\end{equation}
The estimated values range between a few $ \times 10^{39}$ and $10^{40}$ \ergs, again reflecting the range of column densities.

The energy associated with the nuclear starburst as derived by \citet{veille94} is estimated to be $\sim $$6 \times 10^{42} {n}_{\rm e}^{-1}$ \ergs. 
\citet{veille94} quote that for values of ${n}_{\rm e} \geq 40$ cm$^{-3}$, the injection rate of the nuclear starburst would have been able to produce the \HI\ outflow (i.e.\ providing a power of $> 10^{41}$ \ergs). 

Recent results (e.g.\ Nyland et al.\ 2013) have shown that fast \HI\ (and molecular gas) outflows can also be driven by low-power radio sources due to the  high efficiency of the coupling between the radio plasma with the surrounding ISM/IGM  e.g.\ \citet{Cavagnolo2010}. Thus, it is worth exploring  the energetics connected to the radio plasma also for NGC~3079.
Indeed, the presence of an interaction of the parsec-scale jet with ambient gas in NGC~3079 has been shown by \citet{Cecil2001} and \citet{middel07}. 

A number of studies have recently addressed the crucial issue of the conversion between radio luminosities and jet power. We use the relations proposed by 
\cite{willo99}, \cite{wu09}, \citet{Birzan2008} and \citet{Cavagnolo2010}.
Despite the different approaches used  by these authors in deriving these relations, we obtain a reasonable agreement from these different methods, indicating a jet power of  $\sim$$ 4 \times 10^{41}$ \ergs.
Thus, also the jet power appears to be large enough to drive  the \HI\ outflow. This is further supported by the presence of  jet-cloud interactions in the galaxy, as suggested by \citet{Cecil2001} and \citet{middel07}. 

However, a caveat to the radio plasma driving the outflow is whether the putative VLBI jet is really connected to the radio lobes. 
Interestingly, the position angle of the VLBI scale radio jet is close to the plane of the galaxy \citep{irwin88}, and misaligned by about $65^{\circ}$ with respect to the large-scale lobes. In this situation, one could hypothesise that  this is due to a deflection  partly caused by the interaction with the medium: this would be supported by the presence of broad blue-shifted \HI\ absorption. A similar (albeit  not so extreme) misalignment was found in at least an other spiral galaxy - IC~5063 -  where the jet is seen in the plane of the galaxy. Interestingly,  IC~5063 also hosts massive outflows of ionised, \HI\ and molecular gas (\citealt{morga07}, 2015; \citealt{Tadhunter2014}). 

Thus, our considerations show that it is plausible for the radio jet to have a role in driving the \HI\ outflow. However, the connection between the complex radio continuum structure in the inner pc region \citep{trot98,swada00,hagiwa04,kondra05,middel07} and the large-scale disk, and also the connection between the pc \citep{kondra05} and kpc-scale outflows in this galaxy still remains ambigiuous.

\section{Conclusions}

We have presented deep \HI\ observations of the famous starburst/Seyfert galaxy NGC~3079 and its surroundings. The improved sensitivity of the data compared to those in the literature have allowed to uncover  new and  interesting features which give an improved picture of the hectic evolution of NGC 3079 and its environment: 

\begin{enumerate}
 
\item NGC~3079 has a more extended (out to $\sim$22 kpc radius) and disturbed \HI\ disk  than previously observed. The \HI\ disk is lopsided on all scales, both morphologically and kinematically, and shows a mild warp on both the north and south sides. The strong  kinematical lopsidedness in the central regions, and the non-circular motions associated with that, are possibly connected to the high levels of star formation in the central regions.

\item  NGC 3079 is clearly interacting with its neighbours. Large streams of gas and tails are visible in the NGC~3079 group. The  disk of NGC 3079 is strongly affected by these interactions.

\item In addition to the two previously known companion galaxies (NGC 3073 and MCG~9-17-9), \HI\ gas has been detected from five more galaxies (three with previously unknown redshift) which are likely associated with the NGC~3079 group. Among these, two of them ([YGK81] 078 and SDSS~J100311.18+553557.6) seem to be taking part in the interactions in the group.  

\item A long \HI\ bridge is visible between NGC 3079 and MCG 9-17-9 spanning over a projected length of $\sim$33 kpc. We suggest that ram pressure due to the hot X-ray halo of NGC~3079 is most likely responsible for the stripping of gas from MCG~9-17-9, although also tidal effects may have  been relevant in the past.

\item The cometary \HI\ tail of NGC 3073, earlier discovered by \citet{b1}, extends even further, while a second, shorter tail is now also revealed. Our result agrees with the conclusion of \citet{b1} that  ram pressure by the super wind coming from NGC 3079 is responsible for this tail. 

\item An \HI\ counterpart to the ionised gas outflow of NGC~3079 has been detected in the form of a blueshifted wing of the \HI\ absorption profile against the central continuum source, extending to velocities well outside the range of rotational velocities.  However, we cannot distinguish between the starburst and AGN driving this outflow, with both having comparable power sufficient to drive the outflow.
 
\end{enumerate}

\section{Acknowledgements} 
The WSRT is operated by ASTRON, the Netherlands Institute for Radio Astronomy, with the support from the Netherlands Foundation
for Scientific Research (NWO). N.S.\ acknowledges support by the Square Kilometre Array (SKA) and the Hartebeesthoek Radio Astronomy Observatory (HartRAO). S.C.\ acknowledges support by the South African Research Chairs Initiative of the Department of Science and Technology and National Research Foundation and SKA. 
R.M.\ gratefully acknowledge support from the European Research Council under the European Union's Seventh Framework Programme (FP/2007-2013)/ERC Advanced Grant RADIOLIFE-320745.

\appendix

\section{Channel maps}
\begin{figure*}
\centering
\includegraphics[width=17cm]{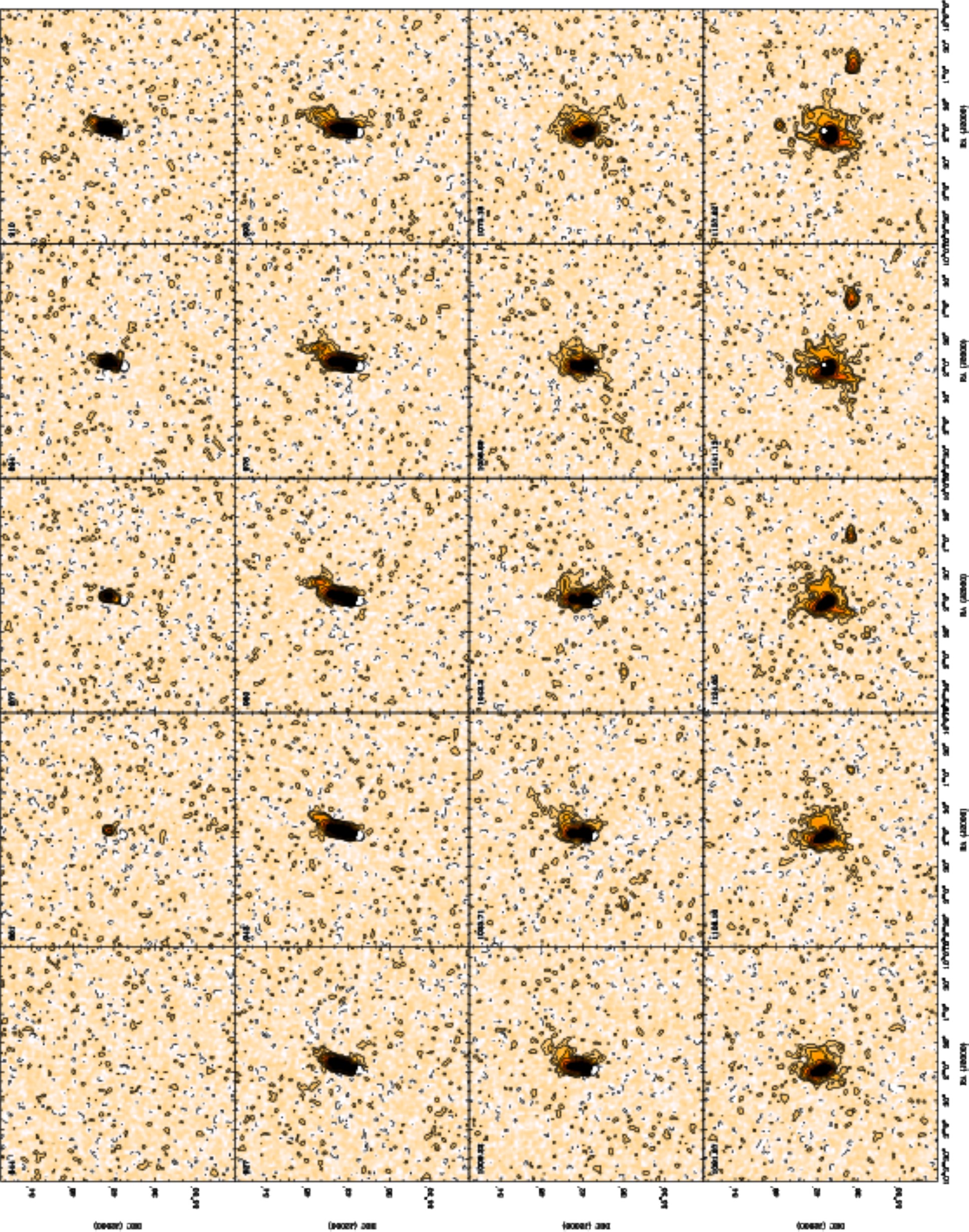}
\caption{Selected  channel maps at $45^{\prime\prime} \times 48^{\prime\prime}$ resolution, in every second channel, containing \HI\ emission. Lowest positive  contour level    0.3 \mJybeam\ which is also the step in contour level. The negative contour is --0.3 \mJybeam. The channel  velocity (in \kms) is marked at the top left corner of each panel. }
\label{f:chan}
\end{figure*}

\begin{figure*}
\centering
\includegraphics[width=17cm]{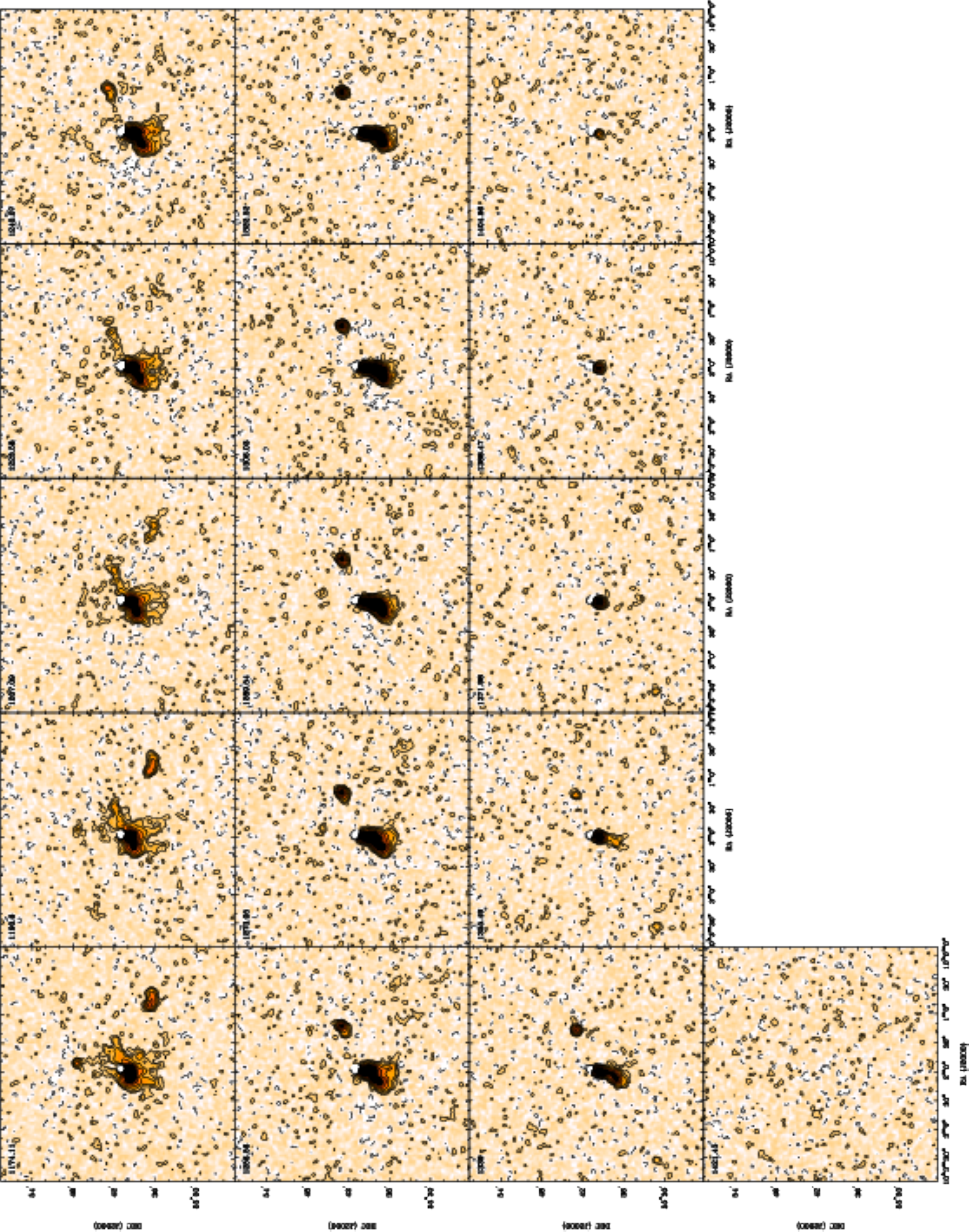}
\caption[]{Fig.\ \ref{f:chan} continued.}
\end{figure*}

\label{lastpage}

\end{document}